\documentclass[traditabstract]{aa}
\usepackage{natbib}
\usepackage{graphicx}
\bibpunct{(}{)}{;}{a}{}{,}

\newcommand{\ks}{km~s$^{-1}$}
\newcommand{\nn}{\accent'27A}
\newcommand{\ms}{M$_{\odot}$}
\newcommand{\rs}{R$_{\odot}$}
\newcommand{\vs}{KX~Vel}

\newcommand{\ANG}{\accent'27A}

\newcommand{\he}{\ion{He}{i}~6678~\ANG\ }

\begin{document}

\title{New and revised parameters for several southern OB binaries
\thanks{Based on data products from observations made with ESO Telescopes
at the La Silla Paranal Observatory under program ID: 074.D-0300(A),
075.D-0061(A), 075.D-0103(A), 077.B-0348(A), 077.D-0321(A), 077.D-0635(A),
079.D-0564(A), 079.D-0564(B), 081.D-2008(A), 081.D-2008(B), 083.D-0589(A),
086.D-0997(A), 087.D-0946(A), 089.D-0189(A), and 089.D-0975(A).}}

\author{Pavel Mayer\inst{1}\and Horst Drechsel\inst{2}\and Andreas Irrgang\inst{2}}
\institute{Astronomical Institute of the Charles University, Faculty of Mathematics and Physics,
 V Hole\v sovi\v ck\'ach 2, \\180 00 Praha 8, Czech Republic
\and
Dr. Karl-Remeis-Observatory \& ECAP, Astronomical Institute, Friedrich-Alexander University
Erlangen-Nuremberg, Sternwartstr. 7, D-96094 Bamberg, Germany
}

\titlerunning{Several southern OB binaries}


\date{Release \today}

\abstract
{Using ESO FEROS archive spectra of several southern OB-type binaries,
we derived periods for three SB2 spectroscopic binaries, HD~97166,
HD~115455, and HD~123590, and two SB1 systems, HD~130298 and HD~163892. It
was also possible to use new FEROS spectra to improve the parameters of
the known binaries, KX~Vel and HD~167263. For KX~Vel, we determined a dynamic
mass of the primary of 16.8~\ms, while the evolutionary model suggests a higher
value of 20.2~\ms. We derived an improved period for HD~167263, and in its spectra,
we recognized contributions of both of its interferometric components. }


\maketitle

\section{Introduction}
Massive stars of spectral types O and early B deserve special attention because 
these extremely rare objects are the main drivers of the chemical and 
dynamical evolution of galaxies and play an important role as cosmological
probes. They are crucial in the comparison 
to recent refinements of the theory of stellar structure and evolution 
of massive stars with respect to mass loss, core overshooting, rotation, 
and metallicity. The still persisting discrepancy between expected 
and observed masses of very massive stars also needs to be resolved.  
In particular, binaries offer the opportunity to provide precise absolute 
stellar quantities (masses, radii, and luminosities) for challenging
theoretical  predictions. Listings of O-type stars such as those by \citet{mason98}
and \citet{MA04} contain many early-type objects, for which their status as single
or double stars -- either SB1 or SB2 -- is still uncertain.
Recently, \citet{chini} presented an extensive spectroscopic survey of OB
stars regarding their binarity, which was based on several thousand spectra obtained
with the ESO FEROS and Bochum University BESO echelle spectrographs over 
the past years.

The FEROS\footnote{FEROS is an echelle fiber spectrograph used with the
2.2m telescope MPIA/ESO at La Silla and supplies spectra with the
resolving power of 48000 in the region from 3575 to 9215~\ANG.} archive 
of ESO is a rich source of spectroscopic data. It contains a wealth of
O-type star spectra, most of which can be directly used in 
pipeline-reduced form. In the course of ourlong-term project to analyze
early-type binaries and multiple systems, we mainly used FEROS archive 
data to identify hitherto unknown binaries among O stars and to obtain
their fundamental parameters. An example of an already published study
is the paper on HD~165246 \citep{ma13}.

Here, we present additional new results based on the study of the FEROS spectra
of several southern O-type stars. Three of the stars -- HD~97166,
HD~115455, and HD~123590 -- were found to be double-lined spectroscopic
binaries; HD~130298 and HD~163892 could be identified as single-lined 
binaries (Sect. 2). We measured the radial velocities (RVs) using 
Gaussian fits. In a recent paper by \citet{sota13}, approximate 
periods for four of these five binaries were published. Our values were 
found independently of this paper. In Sect. 3, we present a revised 
study of the two OB binaries KX~Vel and HD~167263 and derive improved
parameters based on an analysis of new FEROS spectra.

\section{New radial velocity curves and orbital periods for five OB binaries}
\subsection{HD 97166}
According to \citet{wal73}, HD~97166 is an O7.5\,III((f)) star; \citet{sota} give
a slightly different classification of this binary as O8\,IV (in their Table 2).
With a $V$ magnitude of 7.89, it is the brightest member of the open
cluster NGC~3572 in the Carina region. \citet{chini} have noted the SB2 nature of
this object. The sample of the FEROS spectra that we analyzed is
listed in Table~\ref{97} with the measured RVs. In most spectra, the
primary and secondary lines are strongly blended. Fortunately, the assignment
of lines to both components and the measurement of their RVs is facilitated
by considerably different temperatures of both components:
for example, lines of \ion{Si}{iii} 
originate only in the cooler component, while the contribution of the 
hotter component is strong in \ion{He}{ii} lines, as seen in Fig.~\ref{tri}.
Therefore, we first measured the narrow \ion{Si}{iii} lines 4553 and 4568~\nn\,,
for which an accuracy of about 1~\ks\, can be achieved. In a second step, the
deconvolution of the line blends and the measurement of the line positions of
the hotter component were carried out by keeping the RVs of the cooler
component at their previously determined values.

\begin{figure}
\resizebox{\hsize}{!}{\includegraphics{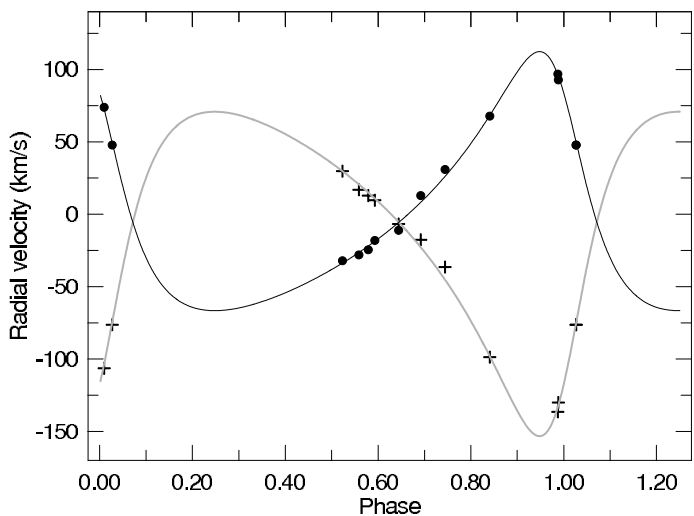}}
\caption{Radial velocity curve of HD~97166 (FEROS spectra); full 
circles -- primary RVs, crosses -- secondary RVs.}
\label{97rv}
\end{figure}

We found only one published RV of this star. It was measured by 
\citet{conti77} and is contained in our Table~1. The error in phase for
this old RV is $\pm 0.06$, which cannot explain the deviation of this
measurement from our RV curve. The use of such an old RV could
potentially improve the accuracy of the period due to its large time
separation, but it would be questionable because of the
severe blending of the primary and secondary lines. It was, therefore, not
included in our data set.

The values of $m_i \sin^3 i$ for the binary components of HD~97166 are large.
The mass of the primary component can be estimated at about 24~\ms\, using a calibration of
O~star parameters by \citet{martins}. Hence, an inclination of
$\approx 70\degr$ can be estimated. Nevertheless, owing to the large separation
of the components, no eclipses or ellipsoidal variations are to be expected
or have so far been mentioned in the literature.

\begin{table}
\caption{Spectra and RVs of HD~97166.}
\label{97}
\begin{tabular}{crrr}
\hline\hline\noalign{\smallskip}
HJD        & RV Pri  &RV Sec&Phase \\
-2400000   & \ks     &\ks   &      \\
\hline\noalign{\smallskip}
42890.5900 &$ -4.0^1$&      &0.6236\\
54211.7307 &  68.0 &  -98.6 &0.8404\\
54599.5973 &$-28.0$&   17.0 &0.5587\\
54600.6240 &$-24.5$&   13.0 &0.5787\\
54601.5756 &$-18.0$&    9.8 &0.5929\\
54625.6280 &  74.0 &$-106.3$&0.0096\\
54626.6215 &  48.0 &$ -76.2$&0.0268\\
54953.6701 &  13.0 &$ -17.6$&0.6916\\
54956.6974 &  31.0 &$ -36.4$&0.7440\\
55605.8678 &  93.0 &$-130.4$&0.9884\\
55643.7316 &$-10.6$&$  -6.5$&0.6442\\
56067.6708 &  97.0 &$-136.4$&0.9873\\
56098.5979 &$-32.0$&  29.8  &0.5230\\
\hline\noalign{\smallskip}
\end{tabular}

Note 1: RV published by \citet{conti77}; only the primary RV is given.
\end{table}

To determine spectral types, we used the $\log W'$ 
criterion by \citet{conti71}, which is based on the ratio $W'$ of 
equivalent widths for the lines \ion{He}{i} 4471~\ANG\, and \ion{He}{ii}
4541~\ANG. The values of $\log W'$ are 0.05 for the primary and $\approx
0.4$ for the secondary components, respectively. The former value agrees
with the original classification of O7.5 by \citet{wal72}, and the
latter value suggests a type O9 for the secondary component, which
is probably of luminosity class V. The luminosity ratio $L_2/L_1$ 
can be estimated from the strength of the \ion{He}{i} line 5876~\ANG, since this line is
nearly independent of temperature and $\log g$ in the possible parameter
range for this system. Accordingly, the ratio $L_2/L_1$ is 0.45, which
corresponds to a magnitude difference of 0\fm85. Therefore, a compromise classification
of the primary as a class O7.5\,IV object is suggested.
Adopting the calibration by \citet{martins}, the absolute magnitudes of the primary
and secondary components are $\approx -5\fm0$ and $-3\fm9$, respectively. With an
integral absolute magnitude of $-5\fm34$, an observed $V$ brightness of $7\fm87$
and an absorption of $A_V=1.28$, we obtain an absolute visual magnitude of
$M_V=-5\fm7$. This implies a distance modulus of $11\fm93$ for HD~97166, which
is close to the value of $12\fm22$ found by \citet{moffat} for the cluster
NGC~3572. Elements of the orbit are contained in Table~\ref{ELE}.

\subsection{HD~115455}
The binary HD~115455 is another O7.5\,III((f)) star \citep{wal73}. It is a member
of the open cluster Stock~16 with $V=7.95$. The RVs of its
primary component were published by \citet[][5 RVs]{crampton}, who noted
the RV variability, and \citet[][1 RV; SL]{stick01}. It was possible to
include these old velocities in our calculation of the period, resulting
in a considerable improvement of its accuracy.

\begin{figure}
\resizebox{\hsize}{!}{\includegraphics{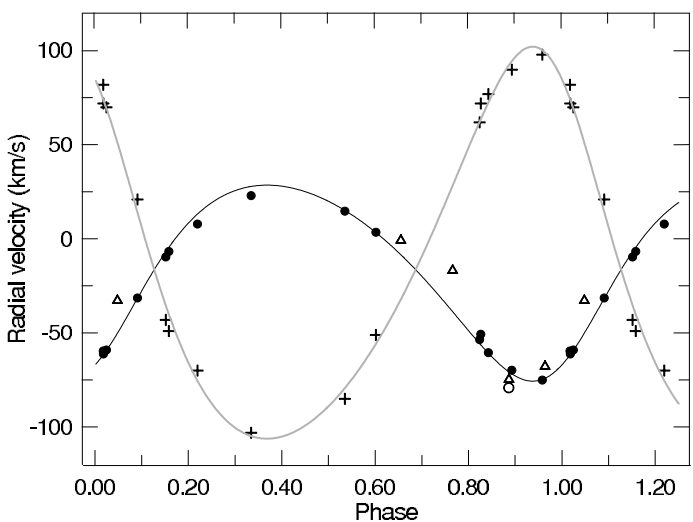}}
\caption{Radial velocity curve of HD~115455. Primary RVs: full
 circles -- FEROS, triangles -- \citet{crampton}, open 
 circle -- SL. Secondary RVs: crosses -- FEROS.}
\label{115}
\end{figure}

The new primary velocities were obtained from the \ion{He}{ii} lines
4541 and 5411~\nn. Since the cooler secondary does not produce a
measurable contribution to these lines, the \ion{He}{i} lines 4922 and
5876~\nn\, were used to determine the RVs of this binary component. All RVs are 
listed in Table~\ref{HD115}, and the RV curve is shown in  Fig.~\ref{115}.
Two deviating RVs by \citet{crampton} might be affected by
the vicinity of the conjunction. Orbital elements are presented in
Table~\ref{ELE}. Examples of two profiles of the \ion{He}{i} 4922~\ANG\, 
line close to the quadrature phases are shown in Fig.~\ref{tri}.

As in the subsection 2.1, $L_2/L_1$ was estimated from the \ion{He}{i} 
line 5876~\ANG. Except for the \ion{He}{i} lines, the primary and 
secondary components can also be discerned in \ion{Mg}{ii} and
\ion{Si}{iii} lines, while \ion{O}{ii} lines can only be attributed to
the secondary component. Although the intensity ratios of various lines
give slightly different results, $L_2/L_1$ equals about 0.10. The 
secondary spectral type can therefore be estimated as B2\,V.

The distance to the cluster Stock~16 was given as $1900\pm 100$~pc by
\citet{vazq}. However, given the expected total magnitude of the binary 
of $M_V=-5.5$ and $V_0=6.4$, the distance would be 2400~pc.

\begin{figure}
\resizebox{\hsize}{!}{\includegraphics{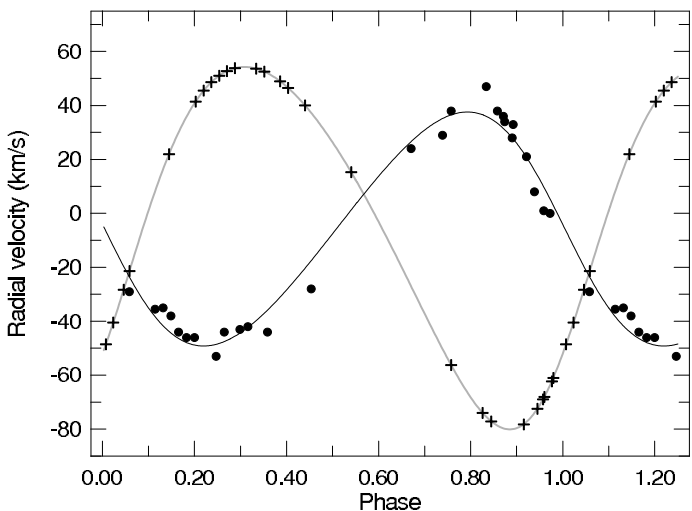}}
\caption{Radial velocity curve of HD~123590. Primary RVs: full
 circles. Secondary RVs: crosses.}
\label{123}
\end{figure}

\begin{figure*}
\begin{tabular}{ccc}
\includegraphics[width=57mm]{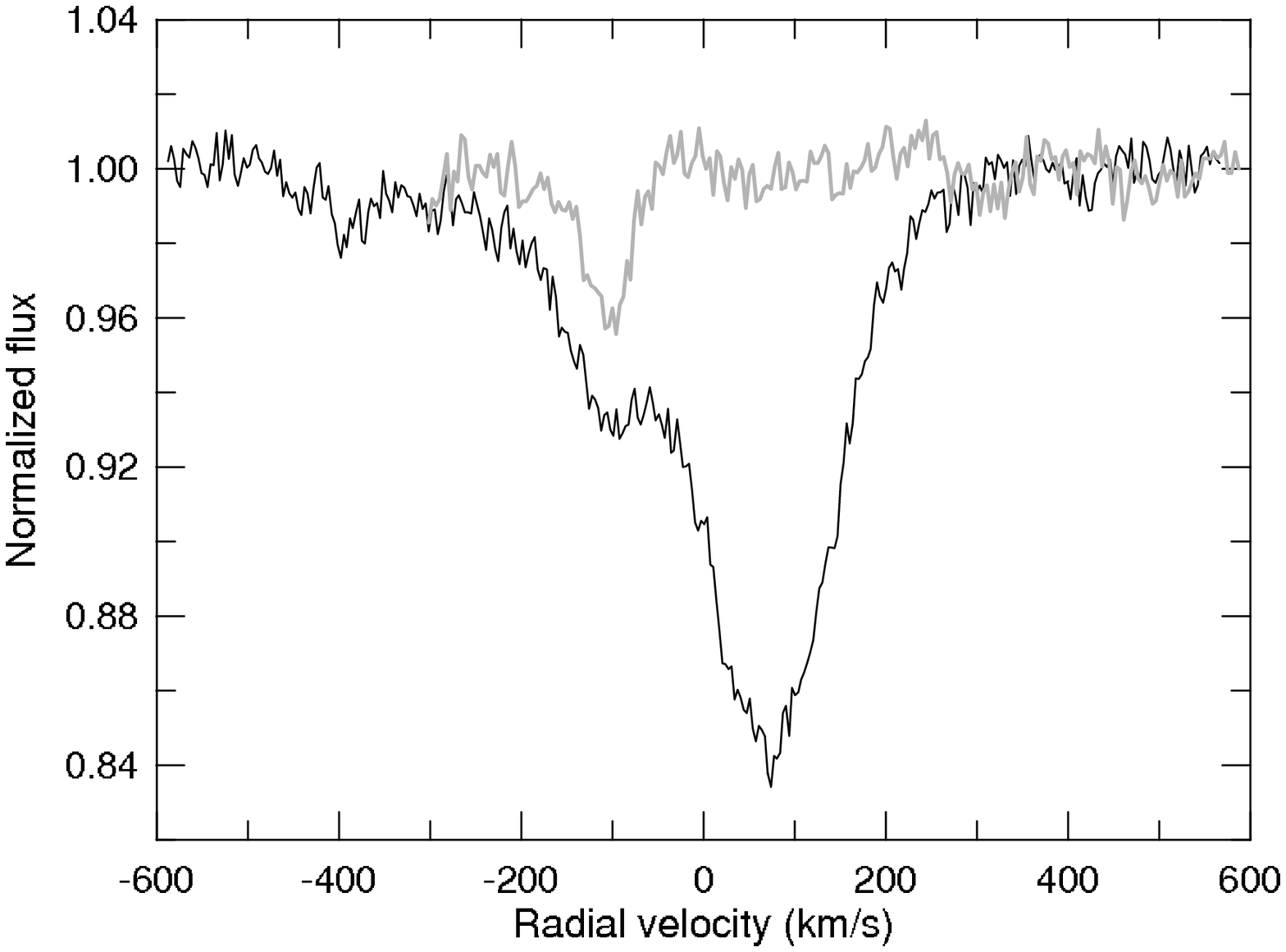}&
\includegraphics[width=57mm]{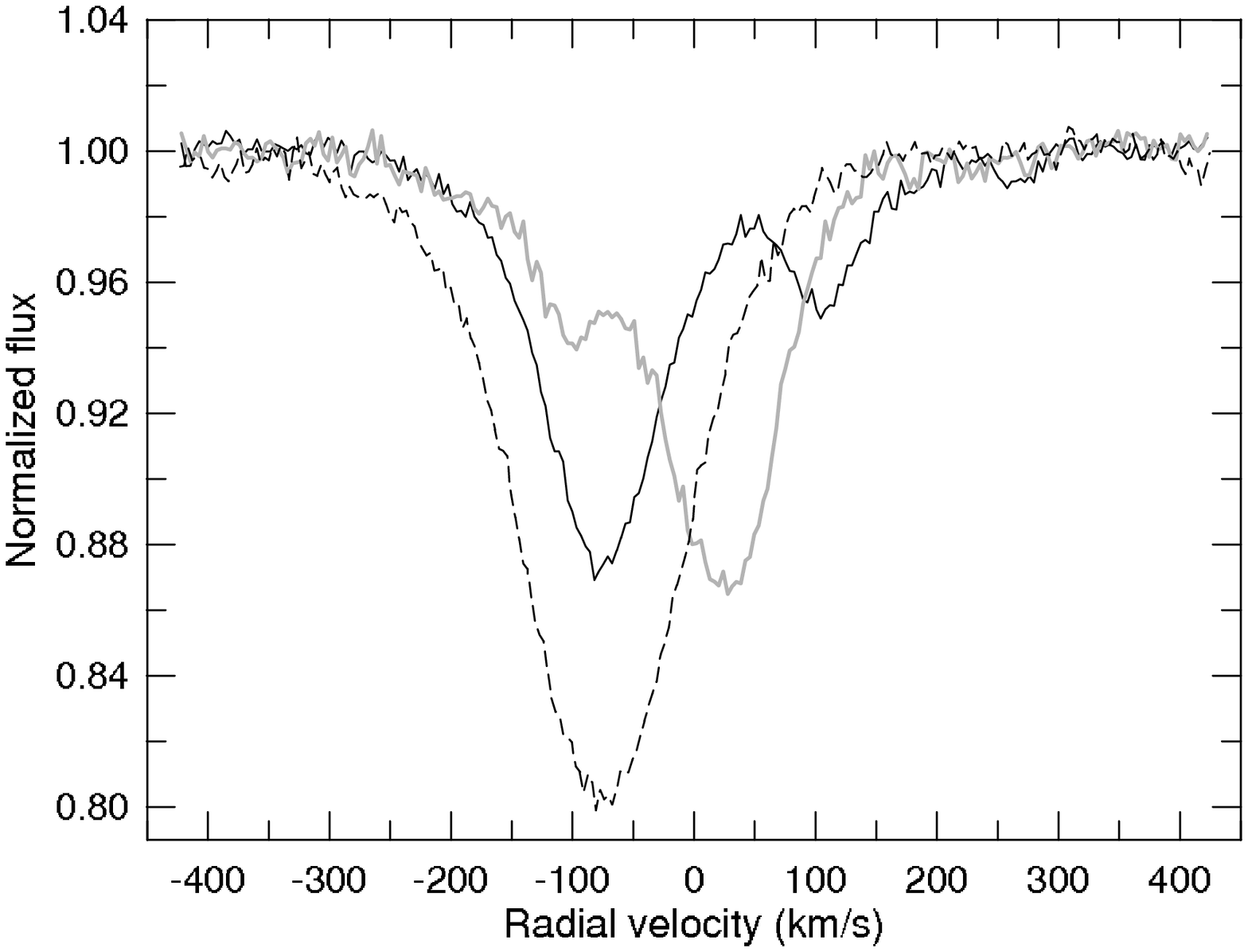}&
\includegraphics[width=57mm]{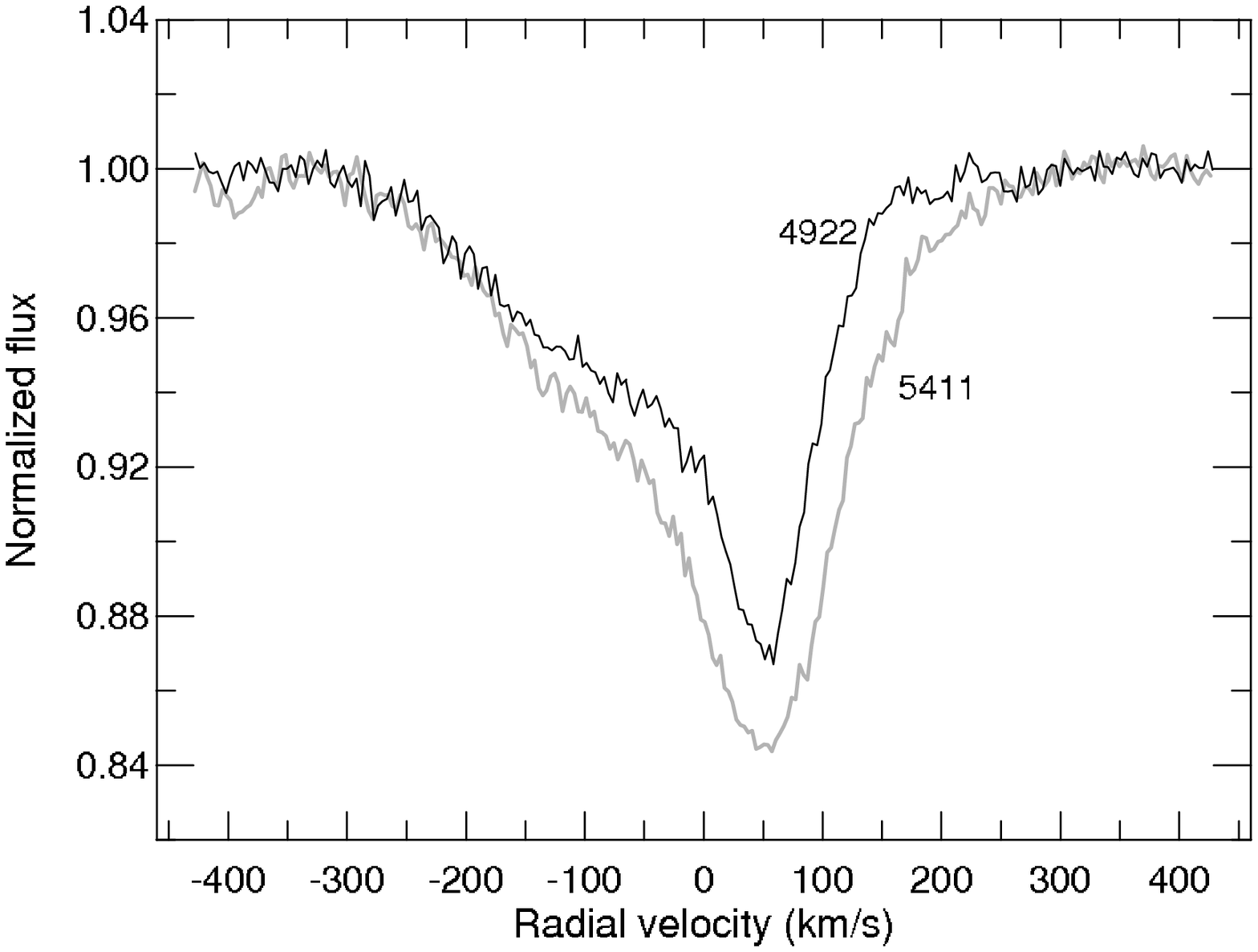}
\end{tabular}
\caption{Examples of line profiles of three stars. Left: HD~97166;
\ion{Si}{iii} 4553~\nn\, (gray) and \ion{He}{ii} 5411~\nn\, (black) in
the spectrum of JD~2454625. Middle: HD~115455; lines \ion{He}{i} 
4922~\nn\, (black) and \ion{He}{ii} 5411~\nn\, (dashed) at JD~2453967 and 
\ion{He}{i} 4922~\nn\, at 2454244 (gray). Right: HD~123590; 
\ion{He}{i} 4922~\nn\, and \ion{He}{ii} 5411~\nn\, at JD~2453970.}
\label{tri}
\end{figure*}

\subsection{HD~123590}
For this star ($V=7.62$), no classification based on a slit-spectrogram 
is available. It is an SB2 system with wide and narrow line components. The 
wide component has the smaller RV amplitude, so it belongs to the primary. We
estimate the spectral type of both components as O7. However, the narrow
component is also present in the \ion{He}{ii} line 6683~\ANG, so its 
luminosity might be higher than that of the wide component. This is also
supported by the width of the \ion{He}{ii} line 5411~\ANG, as seen in 
Fig.~\ref{tri}: it is known that the lower $g$ lines are 
intrinsically wider \citep{LaHu}. Due to the strong blending of the broad 
lines, we did not attempt to measure RVs of \ion{He}{ii} lines.

\begin{table}
\caption{Spectra and RVs of HD~115455.}
\label{HD115}
\begin{tabular}{crrr}
\hline\hline\noalign{\smallskip}
HJD        & RV Pri & RV Sec&Phase \\
-2400000   &  \ks    & \ks  &      \\
\hline\noalign{\smallskip}
53373.8182 &    3.5  &$ -51$&0.6017\\
53965.4890 &$ -50.7$ &   72 &0.8265\\
53966.4905 &$ -69.8$ &   90 &0.8929\\
53967.4806 &$ -75.0$ &   98 &0.9585\\ 
53968.4802 &$ -59.0$ &   70 &0.0248\\
53969.4817 &$ -31.4$ &   21 &0.0912\\ 
53970.4942 &$  -6.7$ &$ -49$&0.1583\\
54191.7093 &$ -53.5$ &   62 &0.8237\\
54209.7216 &$ -59.6$ &   82 &0.0179\\
54209.7350 &$ -61.2$ &   72 &0.0187\\
54211.7461 &$  -9.6$ &$ -43$&0.1521\\
54212.7669 &   7.9   &$ -70$&0.2197\\
54244.6680 &  23.0   &$-103$&0.3346\\
54247.6990 &  14.7   &$ -85$&0.5356\\
54267.5071 &$-60.5$  &   77 &0.8488\\
\hline\noalign{\smallskip}
\end{tabular}
\end{table}

\begin{table}
\caption{Spectra and RVs of HD~123590.}
\label{HD123}
\begin{tabular}{crrr}
\hline\hline\noalign{\smallskip}
HJD        & RV Pri& RV Sec & Phase  \\
-2400000   & \ks   & \ks    &        \\
\hline\noalign{\smallskip}
 53965.4771 &$-35$ &  42.2  & 0.2015 \\
 53966.5065 &$-35$ &  44.5  & 0.2190 \\
 53967.4946 &$-38$ &  48.9  & 0.2357 \\
 53968.4942 &$-44$ &  50.8  & 0.2527 \\
 53969.4958 &$-46$ &  53    & 0.2697 \\
 53970.5140 &$-46$ &  55.3  & 0.2870 \\
 54191.7552 &   1  &$-29.0$ & 0.0454 \\
 54209.7485 &$-44$ &  52.7  & 0.3510 \\
 54211.7601 &$-43$ &  48.8  & 0.3825 \\
 54212.7774 &$-42$ &  45.1  & 0.4025 \\
 54244.6799 &  38  &$-71.9$ & 0.9444 \\
 54246.6991 &  33  &$-62.6$ & 0.9787 \\
 54267.5692 &$-53$ &  54.2  & 0.3332 \\
 54292.4910 &  24  &$-57.1$ & 0.7566 \\
 54296.5114 &  29  &$-73.2$ & 0.8249 \\
 54297.6247 &  38  &$-77.1$ & 0.8438 \\
 54338.5755 &$-28$ &  15.3  & 0.5395 \\
 54480.8590 &  36  &$-68.6$ & 0.9565 \\
 54539.8975 &  34  &$-67.2$ & 0.9594 \\
 54542.6808 &  21  &$-49.7$ & 0.0067 \\
 54545.6954 &   0  &$-22.1$ & 0.0579 \\
 54550.7541 &$-29$ &$ 21.3$ & 0.1438 \\
 54599.7173 &  28  &$-62.8$ & 0.9756 \\
 54955.6960 &   8  &$-40.8$ & 0.0228 \\
 56067.7963 &  47  &$-77.8$ & 0.9145 \\
 56098.6775 &  44  &  40.4  & 0.4391 \\
\hline\noalign{\smallskip}
\end{tabular}
\end{table}

The measured RVs are listed in Table~\ref{HD123}, and the elements of the
orbit are given in Table~\ref{ELE}. Rotational velocities of both 
components differ considerably, as seen in Fig.~\ref{tri}.

\subsection{HD~130298}
\begin{figure}
\resizebox{\hsize}{!}{\includegraphics{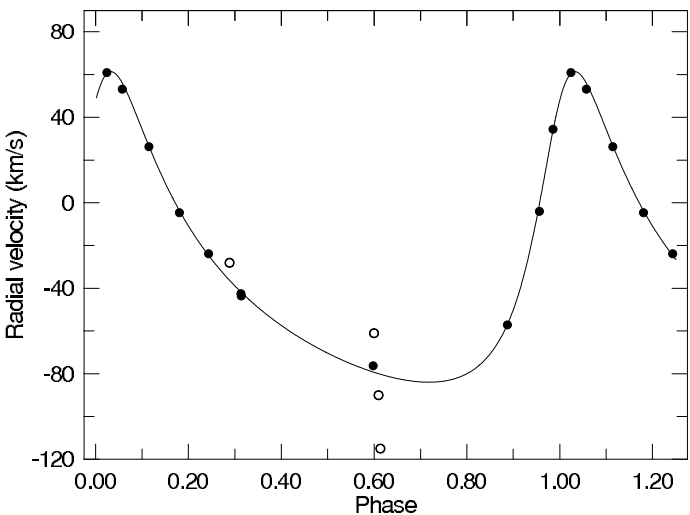}}
\caption{Radial velocity curve of HD~130298. Full circles -- RVs from FEROS spectra,
open circles -- RVs by \citet{feast}.}
\label{RV130}
\end{figure}

This is the earliest binary in our sample with type O6\,III(f)
\citep{GHS} and $V=9.23$. Four RVs were published and their 
variability was noted by \citet{feast}. Although it is listed as an 
emission-line star in SIMBAD, no emission contributions are apparent in
the FEROS spectra in the H$\alpha$ or \ion{He}{II}~4686 \ANG\, profiles. 
\citet{sota13} gave an approximate period of 14\fd63. We refined this
value to $14\fd6302$ and calculated the other elements of the orbit (Table~\ref{ELE}).
The residuals of the FEROS RVs are very small, as documented by 
Fig.~\ref{RV130}. The new period is also compatible with the RVs measured
by \citet{feast}, although the scatter of these measurements is large
(rms=20~\ks). Despite the claim by \citet{chini} that the object
type could be SB2, no traces of the secondary component are evident in the 
\ion{He}{i}, \ion{Si}{iii}, or \ion{O}{ii} lines.

\begin{table}
\caption{Spectra and RVs of HD~130298.}
\label{HD130}
\begin{tabular}{crrr}
\hline\hline\noalign{\smallskip}
HJD        & RV Pri   & Phase  \\
-2400000   &   \ks    &        \\
\hline\noalign{\smallskip}
34578.280  &$-115   $ & 0.6132 & F   \\
34851.501  &$ -28   $ & 0.2882 & F   \\
34973.241  &$ -90   $ & 0.6095 & F   \\
45236.444  &$ -61   $ & 0.5999 & F   \\
54211.7824 &$ -76.2 $ & 0.5977 \\
54246.7136 &   34.5   & 0.9853 \\
54247.7661 &   53.2   & 0.0573 \\
54599.7295 &   26.3   & 0.1146 \\ 
54600.6947 &$  -4.6 $ & 0.1806 \\
54601.6135 &$ -23.8 $ & 0.2434 \\
54625.6620 &$ -57.1 $ & 0.8871 \\
54626.6763 &$  -4.0 $ & 0.9565 \\
54627.6641 &   61.0   & 0.0240 \\
54953.7548 &$ -42.5 $ & 0.3128 \\
54953.7695 &$ -43.6 $ & 0.3139 \\
\hline\noalign{\smallskip}
\end{tabular}

Source: F -- \citet{feast}. The remaining spectra are FEROS spectra.
\end{table}

\begin{figure*}
\begin{tabular}{ccc}
\includegraphics[width=57mm]{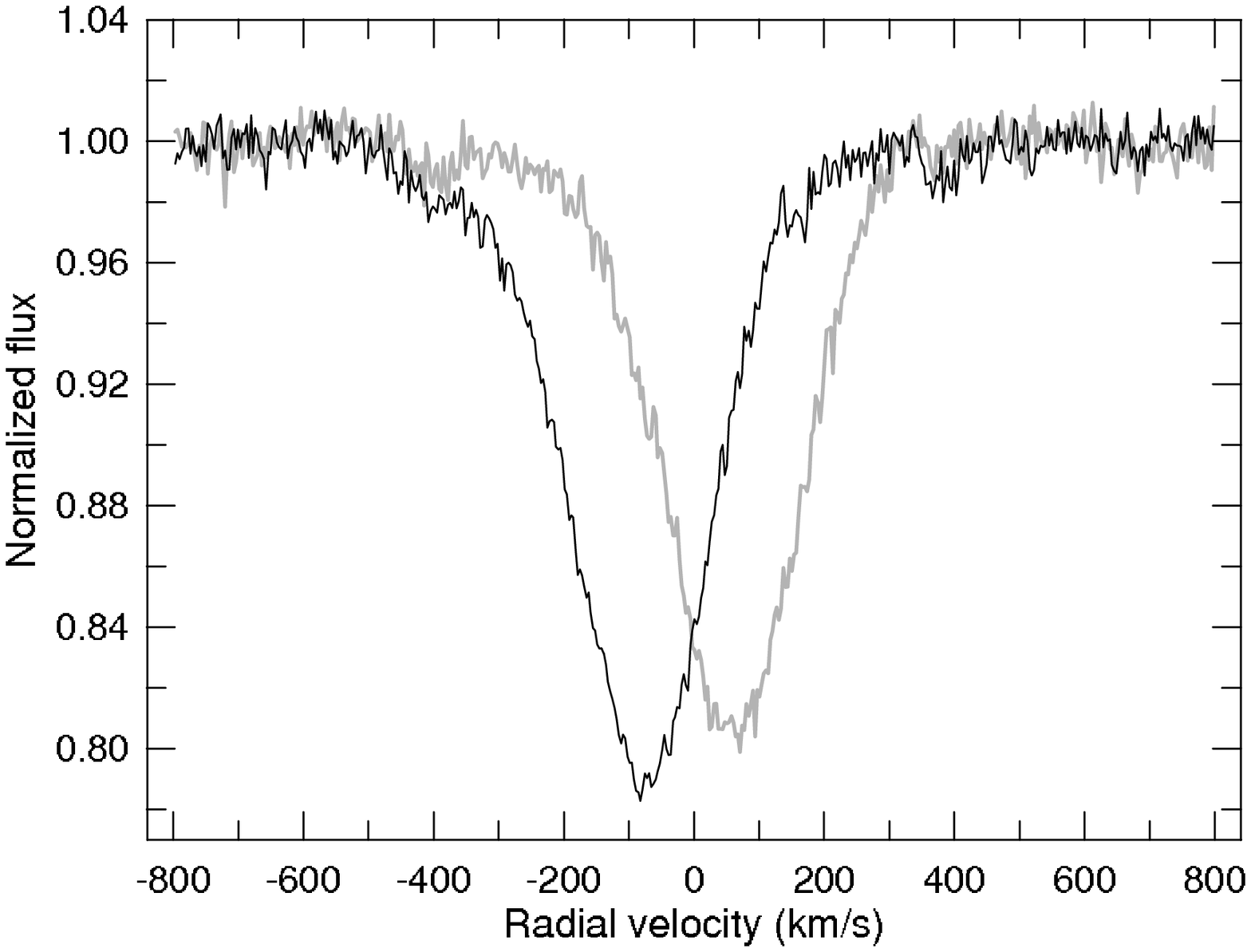}&
\includegraphics[width=57mm]{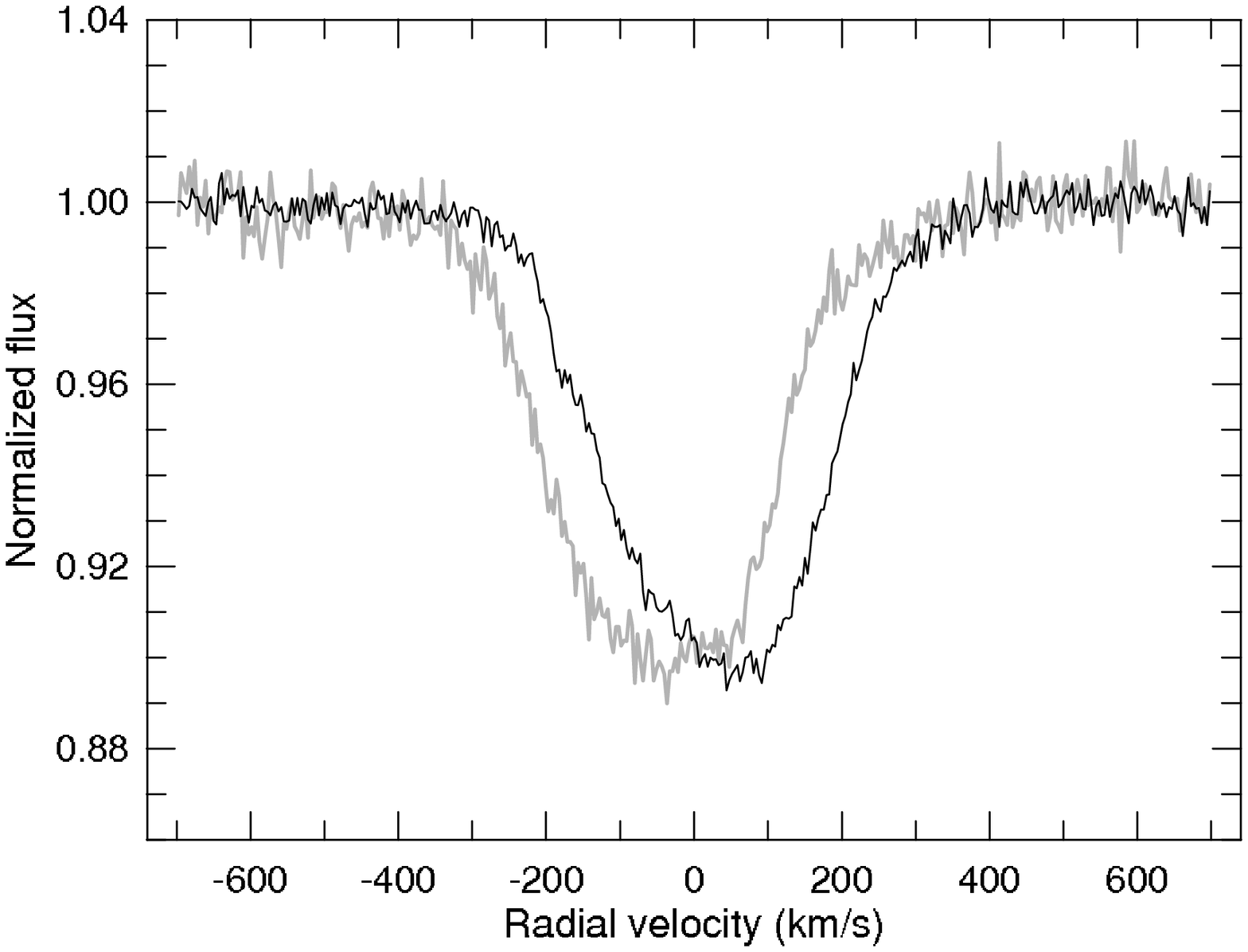}&
\includegraphics[width=57mm]{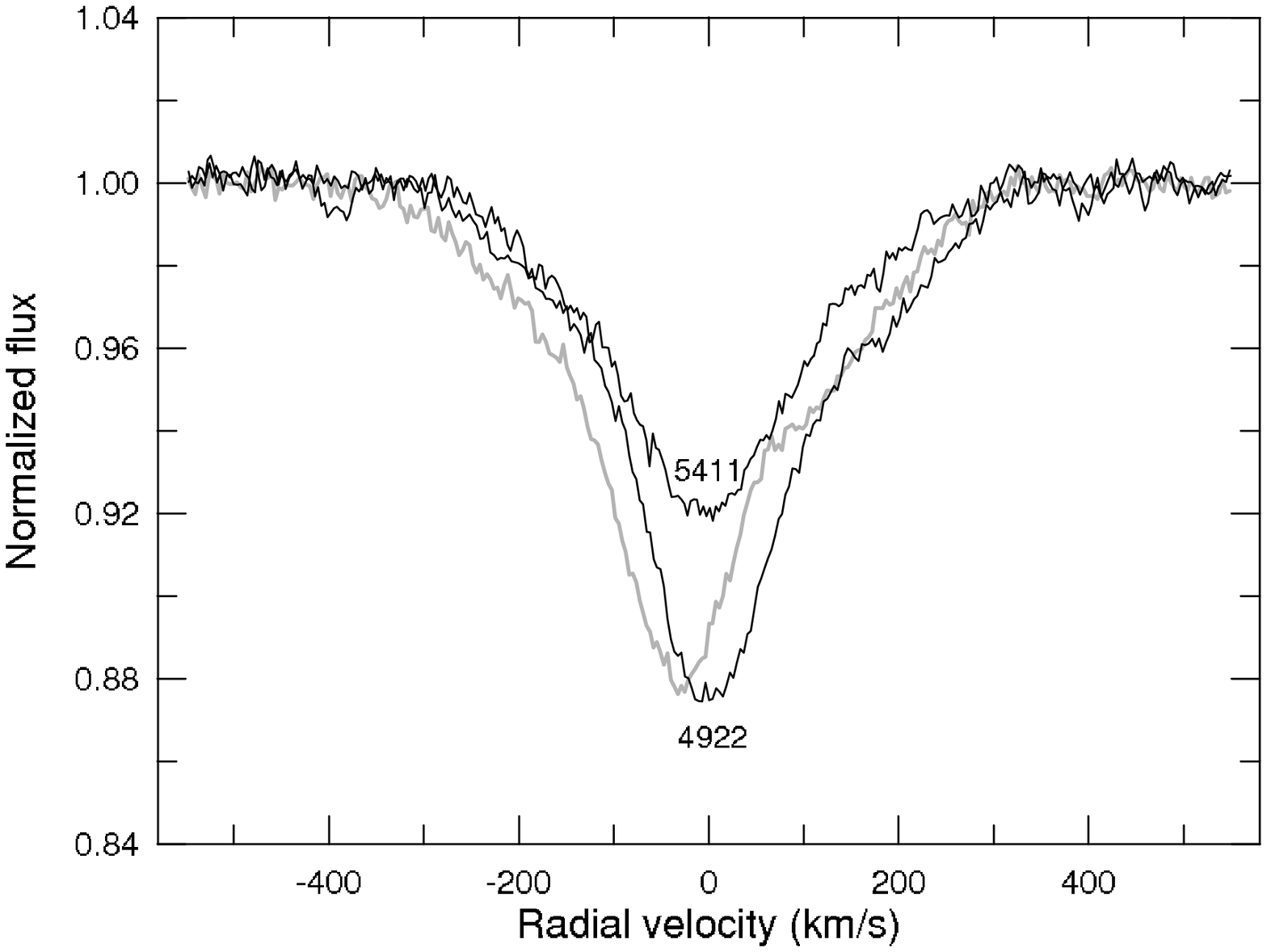}
\end{tabular}
\caption{Examples of profiles of helium lines of three stars. Left:
HD~130298; \ion{He}{ii} line 5411~\nn\, at JD~2454211 (black) and 
2454627 (gray). Middle: HD~163892 \ion{He}{i} line 4922~\nn\, at
JD~2454956 (black) and 2456097 (gray).
Right: HD~167263; spectra of JD~2453483 (only \ion{He}{i} 4922~\nn)
and 2454954 (\ion{He}{i} 4922 and \ion{He}{ii} line 5411~\nn).}
\label{six}
\end{figure*}

\subsection{HD~163892}
\begin{figure}
\resizebox{\hsize}{!}{\includegraphics{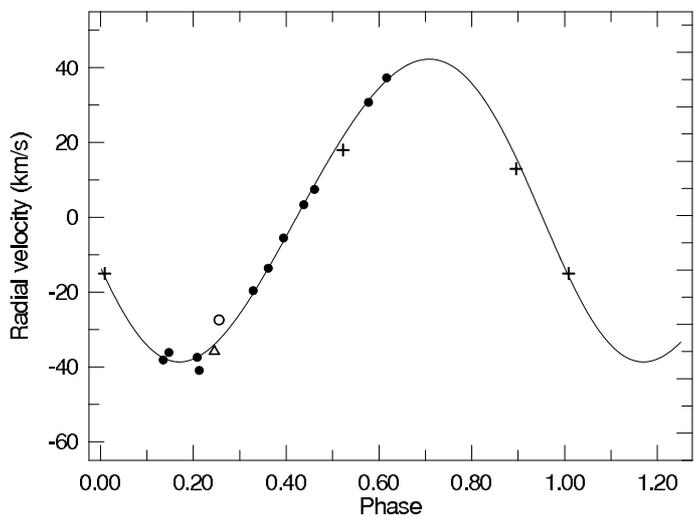}}
\caption{Radial velocity curve of HD~163892. Full circles -- FEROS, 
triangle -- \citet{conti77}, open circle -- SL, crosses -- \citet{feast}. }
\label{1638rv}
\end{figure}

This O9\,IV star \citep{wal73} with brightness $V=7.44$ is a member of the
association Sgr~OB1 \citep{hump}. Several RVs were already published:
\citet{feast}, who noted their variability, \citet{conti77}, and 
SL. These RVs are included in Table~\ref{HD163}. 
Although the RV curve shown in Fig.~\ref{1638rv} appears realistic, it should be
noted that the average RV of the association members is about $-10$~\ks,
but the binary $V\gamma$ differs by $\approx 14$~\ks. The solution of the
orbit is given in Table~\ref{ELE}. We
tried to find any indications for the presence of the secondary 
component but without success. The secondary \ion{He}{i} lines have to 
be at least 20 times fainter than the primary ones.

\begin{table}
\caption{Spectra and RVs of HD~163892.}
\label{HD163}
\begin{tabular}{crrc}
\hline\hline\noalign{\smallskip}
HJD         & RV Pri  & Phase  & Source \\
-2400000    & \ks     &                 \\
\hline\noalign{\smallskip} 
34848.5900  &$ -15$   & 0.0085 & F      \\
34910.3790  &   13:   & 0.8950 & F      \\
35565.5690  &   18    & 0.5220 & F      \\
40044.8300  &$ -35.4 $& 0.2448 & C      \\
46680.8790  &$ -27.4 $& 0.2551 &SL      \\
53546.8814  &   37.3  & 0.6161          \\
54600.8870  &$ -36.1 $& 0.1470          \\
54626.8489  &    7.5  & 0.4607          \\
54953.9244  &$ -37.4 $& 0.2079          \\
54954.8706  &$ -19.6 $& 0.3287          \\
54955.7251  &$   3.4 $& 0.4377          \\
54956.7811  &   30.8  & 0.5725          \\
54976.8538  &$ -38.1 $& 0.1345          \\
56059.8150  &$ -13.6 $& 0.3612          \\
56067.9061  &$  -5.5 $& 0.3940          \\
56097.8446  &$ -40.9 $& 0.2123          \\
\hline\noalign{\smallskip}
\end{tabular}

Source: F -- \citet{feast}, C -- \citet{conti77}, SL -- \citet{stick01}. The
remaining spectra are FEROS spectra.
\end{table}

\begin{table*}
\caption{Elements of the orbits of five binaries with new periods.}
\label{ELE}
\begin{tabular}{lccccc}
\hline\hline\noalign{\smallskip}
Element             &  HD~97166     &  HD~115455    & HD~123590    &HD~130298    & HD~163892    \\
\hline\noalign{\smallskip}
Number of FEROS spectra&    12      &  15           &  26          &  11         &  11          \\
Period [days]       &   57.733(17)  & 15.08410(31)  & 58.867(6)    &14.6302(6)   & 7.834673(33) \\ 
$T_{\rm periastron}$RJD&49140.4(1.5)& 40000.23(39)  & 53894.75(25) &53998.21(3)  & 40003.74(6)  \\
$e$                 &  0.419(18)    &  0.195(17)    & 0.148(4)     &0.469(8)     &  0.065(15)   \\
$\omega$[deg]       & 47(8)         &  213(7)       & 234.3(1.6)   &324.4(1.3)   &112(26)       \\
$K_1$    [\ks]      & 89.5(4.8)     &  52.1(1.4)    & 43.8(7)      &72.8(7)      &40.5(1.5)     \\
$K_2$    [\ks]      & 112.1(3.3)    & 104.3(1.6)    & 67.2(2)      &                            \\
$m_2/m_1$           &0.799(20)      & 0.500         & 0.653(16)    &                            \\
$m_1 \sin^3 i$ [M$_{\odot}$]& 20.4  &  3.76         & 4.89                                      \\
$m_2 \sin^3 i$ [M$_{\odot}$]& 16.3  &  1.88         & 3.19                                      \\
$V\gamma_{\rm pri}$ [\ks]  &$-2.6$(3.4)&$-15.1$(1.0)&$-7.4$(0.8)   &$-38.9$(6)   & 2.8(1.0)     \\
$V\gamma_{\rm sec}$ [\ks]  &$-9.1$(4.3)&$-19.0$(1.9)&$-7.1$(0.7)                                \\
$a \sin i$  [R$_{\odot}$]  &212     & 49.0          &  80.5                                     \\
rms [\ks]           &2.8(pri) 3.5(sec)  & 3.6(pri) 5.4(sec) &4.1(pri) 0.7(sec) &1.5(FEROS)   & 1.8          \\
\hline\noalign{\smallskip}
\end{tabular}
\end{table*}

\section{New orbital elements for two binaries}
\subsection{KX Vel}
The number of eclipsing binaries with an evolved early-type component is 
rather small. For such systems, direct and accurate determinations of 
their masses are important but have rarely been reported. The binary KX~Vel 
(HR~3527, HD~75821, HIP~43413, $V=5.10$) is of type B0\,III, according to
\citet{MCW}, and belongs to this 
group of OB-type systems. \citet{balona} discovered that this star is an
eclipsing variable; hence, a determination of the component masses is in 
principle possible. Electronic RVs of the star were published by
\citet[][MLD]{may97} and by SL. However, the secondary
velocities were determined at only three phases; therefore, the secondary 
semi-amplitude $K_2$ was rather uncertain. Now several more recent 
spectra obtained with FEROS are available. Although the secondary
velocities could still only be measured at a few phases, a better
determination of the secondary semi-amplitude $K_2$ is now possible.

\subsubsection{Spectroscopy}
The FEROS spectra are listed in Table~\ref{FEROS}. Examples of the
spectra taken near quadratures are shown in Fig.~\ref{6678_2}. In 
Table~\ref{FEROS}, the average values of velocities obtained for the 
\ion{He}{i} lines 4713, 4922, 5876, and 6678~\nn\, are listed.
A simplex-based parameter optimization procedure was used to adjust the
orbital elements by a simultaneous fit of the primary and secondary RVs.  
The MLD had also used two measurements of the secondary line \ion{He}{i} 4713~\nn\,
at phases 0.13 and 0.17, but now we consider the line too weak and the RVs
are certainly affected by the vicinity of the primary line. These two RVs
were therefore not included in our RV sample for the
parameter determination. All RVs measured on electronic spectra are plotted in
Fig.~\ref{RV}. The orbital elements can be found in Table~\ref{KX}.

The measured FWHM widths can be used to determine rotational velocities
$V_{\rm rot}\sin i$. For instance, \citet{slettebak} has calibrated the FWHM
versus $V_{\rm rot}\sin i$ relation based on the spectra of a set of early-type
standard stars, and \citet{daflon} also derived such a correlation between the
FWHM of helium lines in synthetic non-LTE spectra and the rotational velocities.
We used a similar calibration by \citet{munari} and the measured FWHM of the
\ion{He}{i} 5876~\nn\, line to derive the rotational velocity
$V_{\rm rot}\sin i$ as $44\pm4$~\ks for both components of KX~Vel.
From Fig.~\ref{6678_2}, it is evident that \ion{He}{i} 6678~\nn\, line profiles
of both binary components can be well represented by Gaussian profiles.

The spectral classification, according to the Yerkes workers
\citep[][Hiltner et al. 1969]{MCW}, is B0\,III, but others give
O9.5\,II, B0\,Ib, or B1\,Ib. However, according to \citet{fraser}, the 
star is not a supergiant. We therefore checked the classification: the
$\log W'$ criterion \citep{conti71}, although used in 
extrapolation, suggests that the type is B\,0.5. To decide about the
luminosity class, the $uvby$ photometry and H$\beta$ index can help.
The measured $\beta$ value is 2.588 \citep{hauck}. According to
\citet{craw}, $\beta$ should be $2.577$ and $2.606$ for the luminosity
classes III and V, respectively. Therefore, the luminosity class of
KX~Vel might be closer to IV than III. Such a luminosity also
agrees with the \ion{He}{i} line 4143~\nn.

\subsubsection{Comparison to synthetic spectra}
\begin{figure}
\resizebox{\hsize}{!}{\includegraphics{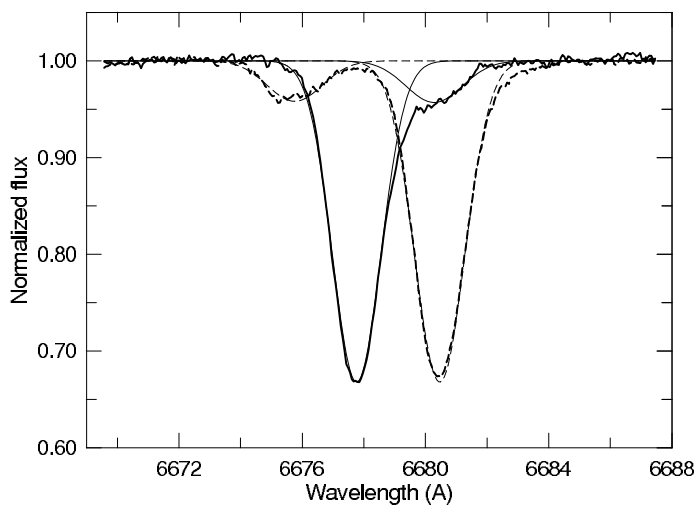}}
\caption{The \he line in two spectra of KX~Vel taken near quadratures 
at JD~2455904 (dashed lines) and JD~2455909 (full lines).}
\label{6678_2}
\end{figure}

The available FEROS spectra of KX~Vel are high in quality (with
S/N exceeding 200) and, hence, allow for a quantitative spectral analysis
with the aim to determine the atmospheric parameters of the binary
components. We applied a new method developed by \citet{irrgang}
for an objective spectroscopic analysis of early-type single and
binary stars. The approach is based on a grid of synthetic spectra
calculated under the assumption of line-blanketed, plane-parallel model
atmospheres with appropriate non-LTE modifications. Using the concept
of $\chi^2$ minimization, the method is able to derive all essential
atmospheric parameters, RVs, and element abundances. It is also
applicable to composite spectra of double-lined spectroscopic binaries,
if the spectra are obtained at orbital phases, where the spectral lines of both
components are not severely blended. We, therefore, used the
near-quadrature spectrum of KX~Vel of JD~2455904 (phase 0.85; the
dashed line in Fig.~\ref{6678_2}).

The composite spectrum is best reproduced by the following parameters:
the primary component has an effective temperature of $T_{{\rm eff},1}
= 29710 \pm 40\mathrm{(stat.)} \pm 600\mathrm{(sys.)} $\,K and the
surface gravity $\log g_1 = 3.669_{-0.004-0.1}^{+0.005+0.1}$; the
secondary values are $T_{{\rm eff},2} = 28070_{-170-870}^{+140+610}$\,K
and $\log g_2 = 4.343_{-0.021-0.434}^{+0.016+0.310}$. The RVs obtained
from the cross-correlation of the complete observed spectrum with its
best-matching synthetic representation are $V_{{\rm rad},1} =
104.0_{-0.1-0.1}^{+0.2+0.1}$\,\ks\, and $V_{{\rm rad},2} =
-110.9_{-1.0-0.4}^{+0.8+0.6}$\,\ks, which are in good agreement with the values
$+104.1$ and $-107.7$\,\ks derived from our Gaussian fit of four \ion{He}{i}
lines (see Table~\ref{FEROS}). The ratio of the
secondary to primary component effective surface areas $A_{\mathrm{eff,s}}/A_{\mathrm{eff,p}}$,
is $0.218_{-0.002-0.011}^{+0.003+0.014}$. This fit parameter weighs
the flux contributions of both stars to the composite spectrum and
assumes a constant surface brightness over
the stellar disks. Hence, it is not identical to the actual geometric
ratio of surface areas because, for example the effect of limb darkening is not
included. However, by taking the ratio of effective surface areas, these
shortcomings might cancel out, so that $A_{\mathrm{eff,s}}/A_{\mathrm{eff,p}}$
is supposed to be very close to the true geometric ratio.

Single-star evolutionary tracks by \citet{ekstrom} are used to
determine the stellar masses ($M_{1} = 20.2_{-1.5\mathrm{(sys.)}}^{+1.8
\mathrm{(sys.)}}\,M_{\sun}$, $M_{2} = 11.5_{-0.5}^{+2.0}\,M_{\sun}$),
ages ($\tau_{1} = 7_{-1}^{+1}\,\mathrm{Myr}$, $\tau_{2} = 0_{-0}^{+10}\,
\mathrm{Myr}$), luminosities ($\log L_1/L_{\sun} = 4.92 \pm 0.12$,
$\log L_2/L_{\sun} = 3.93_{-0.07}^{+0.42}$), and radii
($r_{1} = 10.9_{-1.6}^{+1.7}\,R_{\sun}$, $r_{2} = 3.9_{-0.1}^{+2.9}\,R_{\sun}$)
from the position of the stars in the $(T_{\mathrm{eff}}, \log g)$ diagram.

\subsubsection{Impact of the eclipse light curve} 
It is a remarkable and an unfortunate fact that the photometric coverage of
the eclipse is still limited to the observations by \citet{balona} for more than 20 years
after the discovery of the eclipsing nature. 
Nevertheless, based on a more accurate ephemeris and better knowledge of
the orbit, we tried to model the observed part of the eclipse using the 
code FOTEL \citep{had04}. We read the KX~Vel magnitudes from Fig.~1
of the \citet{balona} paper. The phase of the minimum is accurately
known from spectroscopy; therefore, the 
incomplete light curve defines the shape of the minimum.

However, the partial coverage of the eclipse does not allow for an
unbiased determination of the light curve parameters. It is namely
the width of the eclipse, which is defined by the observations and which
depends mostly on $r_1+r_2$ and inclination. So, we fitted the 
observations by adjusting the parameters $r_1+r_2$ and the inclination
to achieve the best fit quality as judged by the $\chi^2$ sum. Hereby,
we assumed a fixed ratio $r_2/r_1= 0.467$ and luminosity ratio 
$L_{\rm V,2}/L_{\rm V,1}=0.2355$, which resulted from the 
spectroscopic analysis described in the previous section.
With the mentioned constraints, we obtained $r_1+r_2 = 0.167\pm 0.003$
(i.e. $r_1 = 0.114\pm 0.004$  and $r_2=0.053\pm0.002$) and 
$i = 74\fdg 6\pm 0\fdg 7$. The parameter errors were assesed by
confining the range of possible values of the $r_2/r_1$ ratio
used for the light curve solution. The respective representation of the
light curve is shown in Fig.~\ref{exper}, and the corresponding absolute
masses and radii are listed in Table~\ref{ELEM}. The semimajor axis is 
110.7\,(2.2) ~\rs.

\begin{figure}
\resizebox{\hsize}{!}{\includegraphics{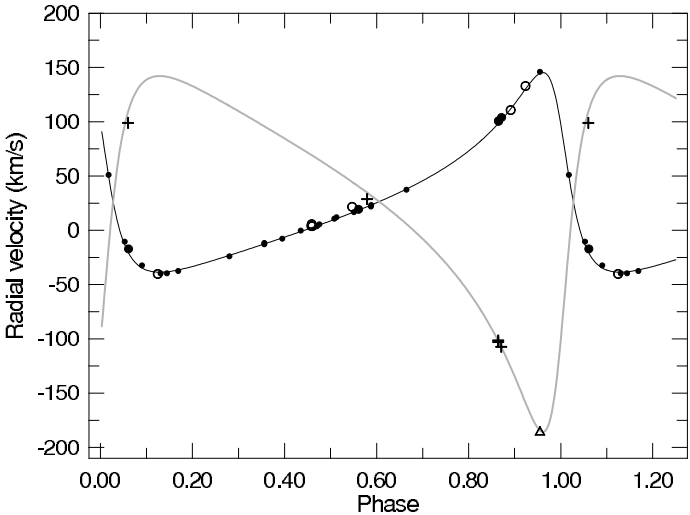}}
\caption{Radial velocities of KX~Vel. Primary RVs: full circles -- FEROS,
open circles -- SL, small full circles -- MLD. Secondary RVs: crosses -- FEROS,
triangle -- MLD.}
\label{RV}
\end{figure}

\begin{figure}
\resizebox{\hsize}{!}{\includegraphics{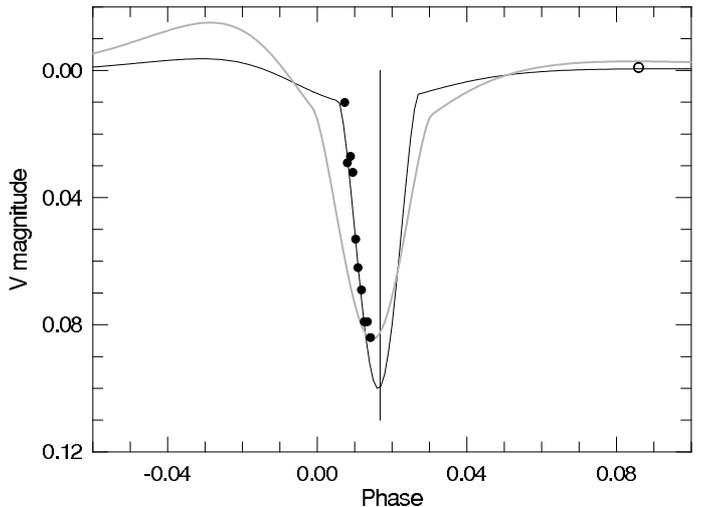}}
\caption{KX~Vel photometry by \citet{balona}. The open circle at the
right represents the level of the measurements out of minimum. The black
curve corresponds to the solution with $i=74.6\degr$; the gray curve to
the solution with $i=60\degr$. The vertical line is drawn at the phase
of the conjunction according to the FOTEL solution of the spectroscopy.
The phase is calculated from the time of periastron. }
\label{exper}
\end{figure}

The primary mass comes out as 16.8~M$_{\odot}$. It must be noted that
the difference of the primary evolutionary and dynamical masses is 
considerable. There are two possibilities to end up with a larger 
dynamical mass -- either the RV amplitude had to be larger, which can be
ruled out by our RV curve (see Fig.~\ref{RV}), or the orbital 
inclination had to be as low as $\approx~60\degr$. In this case no 
eclipses would occur, if the radii of our solution are assumed. At such
a low inclination, the observed eclipse depth could only be reproduced by
radii, which are 56\,\% larger than those following from our analysis. 
As illustrated in Fig.~\ref{exper}, the corresponding light curve (gray
line) is far from being an acceptable representation of the 
observations. Hence, it must be concluded that the difference between the
dynamical and evolutionary masses is real.

\begin{table}
\caption{FEROS spectra of KX~Vel.}
\label{FEROS}
\begin{tabular}{crrrr}
\hline\hline\noalign{\smallskip}
HJD        & Exp.time & Phase  & RV Pri & RV Sec  \\
-2400000   &  s       &        & \ks    & \ks     \\
\hline\noalign{\smallskip}
53371.8581 &  90      & 0.5789 &  19.5  &  28.2   \\
53484.5784 & 150      & 0.8639 & 101.1  &$-102.8$ \\
53484.5808 & 150      & 0.8640 & 100.6  &$-101.3$ \\
55904.8770 & 120      & 0.8703 & 104.1  &$-107.7$ \\
55909.8645 & 120      & 0.0599 &$-17.1$ &  99.0   \\
\hline\noalign{\smallskip}
\end{tabular}
\end{table}

\begin{table}
\caption{Elements of the orbit of KX~Vel.}
\label{KX}
\begin{tabular}{lcccc}
\hline\hline\noalign{\smallskip}
Inclination [deg]   &  74.6\,(7)         \\
Period [days]       &  26.30575\,(13)    \\
$T_{\rm periastron}$& 49147.711\,(22)    \\
$e$                 & 0.598\,(6)         \\
$\omega$[deg]       & 60.3\,(4)          \\
$K_1$    [\ks]      & 92.8\,(9)          \\ 
$K_2$    [\ks]      & 163.5\,(2.7)       \\
$m_1 \sin^3 i$[M$_{\odot}$]& 15.1\,(4)   \\
$m_2 \sin^3 i$[M$_{\odot}$]&  8.6\,(2)   \\
$V\gamma_{\rm pri}$ [\ks]&   27.1\,(4)   \\
$a\sin i$[R$_{\odot}$]&     106.8\,(2.1) \\
\hline\noalign{\smallskip}
\end{tabular}
\end{table}

With the new radial velocities, the earlier published ephemeris changes
only little. The primary and secondary minimum phases of this eccentric
system should be 0.0170 and 0.7485, respectively. However, for the given
parameters, no secondary eclipse can be expected nor is it actually 
evident from the photometric observations. 

\subsubsection{Interstellar lines}
\begin{figure}
\resizebox{\hsize}{!}{\includegraphics{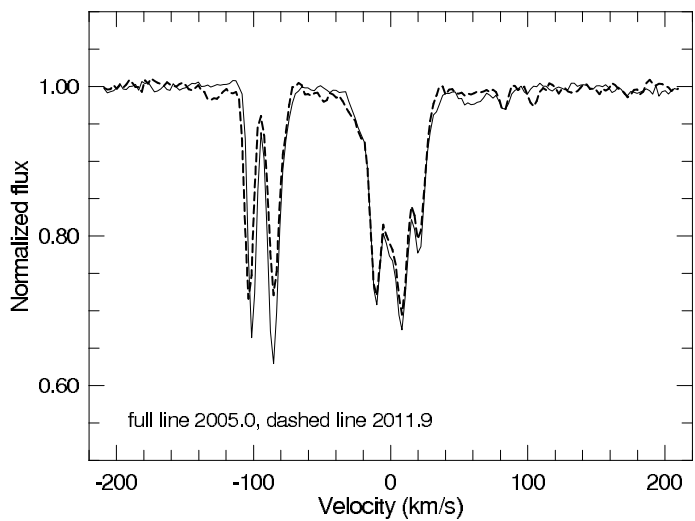}}
\caption{The profile of the interstellar line \ion{Ca}{ii}~3933.66 in the 
spectrum of KX~Vel; the velocity scale is relative to $V_{\rm LSR}$.}
\label{ISL}
\end{figure}

The binary \vs\, is located in the field of the Vela OB~1 association, for which
\citet{hump} gives a distance modulus of 11.37, while KX~Vel is
only at about 10.0. Therefore, the star must be a foreground object. However, it is
actually located behind the Vela supernova remnant and, therefore, 
displays a rich set of interstellar lines: at least five interstellar 
clouds can be recognized in the \ion{Ca}{ii} lines 3934 and 3968~\nn,
of which three are also apparent in the \ion{Na}{ii} lines. Interstellar
\ion{Ca}{ii} and \ion{Na}{ii} lines in \vs\, spectra were studied by 
\citet{danks} (see Fig.~2 of their paper) and by \citet{sembach}, with 
profiles shown in their Fig.~10. For these lines, \citet{sembach} found
changes of their profiles and velocities. With the new spectra, we can
confirm changes of all interstellar features. In Fig.~\ref{ISL}, two 
spectra with a time difference of seven years are superimposed
(to be consistent with Cha \& Sembach, a correction of $-13.1$~\ks\, to
LSR was applied). Although the changes of the velocities are only 
minor (the largest one being observed for the component now at 
$-105$~\ks, while it was at $-98$~\ks\, in the year 1993), changes of the
profiles, namely, the decrease of the equivalent width of the component
with RV $-88$~\ks, are considerable.

\begin{figure}
\resizebox{\hsize}{!}{\includegraphics{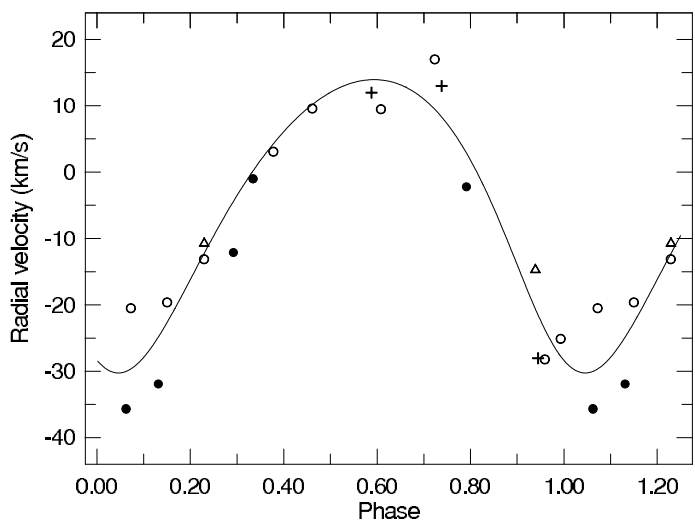}}
\caption{Radial velocity curve of HD~167263; full circles -- FEROS spectra,
crosses -- \citet{feast1}, open circles -- \citet{garmany}, triangles -- SL.}
\label{RV167}
\end{figure}

\begin{table}
\caption{Parameters of the binary KX~Vel.}
\label{ELEM}
\begin{tabular}{lcc}
\hline\hline\noalign{\smallskip}
Parameter    & Primary  & Secondary   \\ 
\hline\hline\noalign{\smallskip}
Mass [\ms]   &  16.8(5)     &  9.5(3) \\
Radius[\rs]  &  12.6(5)     &  5.9(3) \\
$\log g$     & 3.46(3)      &  3.87(3)\\
\hline\noalign{\smallskip}
\end{tabular}
\end{table}

\begin{table}
\caption{FEROS spectra and RVs of HD~167263.}
\label{167RV}
\begin{tabular}{llrrrrrrr}
\hline\hline\noalign{\smallskip}
HJD        & Exp.   & Narrow line & Phase  \\
-2400000   & time s &  RVs \ks    &        \\
\hline\noalign{\smallskip}
53482.9206 & 300    &$-35.6$      & 0.0620 \\
53482.9247 & 300    &$-35.7$      & 0.0623 \\
53855.9301 & 400    &$-12.1$      & 0.2919 \\
54953.9327 & 300    &$-1.0$       & 0.3339 \\
54953.9370 & 300    &$-1.0$       & 0.3342 \\
54976.8684 & 350    &$-31.9$      & 0.1311 \\
55699.9165 & 300    &$-2.2$       & 0.7909 \\
\hline\noalign{\smallskip}
\end{tabular}
\end{table}

\begin{table}
\caption{Elements of the orbit of HD~167263.}
\label{HD}
\begin{tabular}{lcccc}
\hline\hline\noalign{\smallskip}
Period [days]            & 12.76123\,(22)  \\
$T_{\rm periastron}$     & 54005.3\,(7)    \\
$e$                      &     0.181\,(60) \\
$\omega$[deg]            &   156\,(19)     \\
$K_1$    [\ks]           &    22.1\,(2.8)  \\ 
$V\gamma_{\rm pri}$ [\ks]&   $-4.5(1.2)$   \\
rms [\ks]                &            6.1  \\
\hline\noalign{\smallskip}
\end{tabular}
\end{table}

\subsection{HD 167263}
Radial velocities of this binary (16~Sgr, O9.5\,II-III, $V=5.98$) were
published by \citet{feast1}, \citet{garmany} and SL. SL found 
a period of 14\fd75825, noting that a shorter period of 12\fd74 would 
also be possible. This object is a known speckle binary with an
estimated orbital period of 130 years \citep{mason98}. A preliminary
value for the magnitude difference between the two components was 2~mag. 
However, a more recent value of this difference is only
0\fm4\footnote{Fourth Catalog of Interferometric Measurements of Binary
Stars, U.S.Naval Observatory, Washington, D.C.}.
Three published interferometric measurements cover 17.6 years, and the 
position angle changed by $16\fdg3$, so any attempt to calculate the 
visual orbit appears as premature. It seems the orbital period is
probably longer than 130 years.

If the latter small value of the magnitude difference is valid, the
spectral lines of both speckle components should certainly be visible in 
the FEROS spectra. The line profiles consist of a narrow component, which
represents the primary of the spectroscopic binary for which RVs had previously
been derived, and a wide component, which would not be recognized in spectra
of lower resolution. Radial velocities of the narrow parts can be well measured;
RVs originating from various lines were found to differ by less than
$\approx 3$~\ks. The wide part can be measured only approximately; differences
might reach $\approx 10$~\ks, as seen in Fig.~\ref{six}. Very likely, RVs of
the wide profile parts are also changing but not in antiphase with the RVs
of the narrow parts. Therefore, each of the interferometric components might
be a binary itself.

It appears that the narrow profile originates in the fainter 
interferometric component. The spectra and measured RVs are listed in 
Table~\ref{167RV}. The new RVs do not fit the longer period proposed by
SL but are in better agreement with one of their suggested shorter 
periods. The corresponding orbital elements are listed in 
Table~\ref{HD}. The RV curve is rather noisy (Fig.~\ref{RV167}), which might occur 
because the positions of the narrow line components were
affected by the wider ones in spectra of lower dispersion. With the
small number of spectra, their disentangling is uncertain; namely, finding
any difference in spectral types of both components is difficult. 

\section{Conclusions}

We present the results of a spectroscopic analysis of seven OB-type binaries.
New and improved orbital and stellar parameters for these members of
the important and rare class of early-type systems were obtained.
It was demonstrated that a modest number of spectra were
sufficient to find the period and other parameters of these binaries.
With the exception of HD~123590, the number of available RVs for our objects
were relatively low. Additional spectra would be certainly welcome to
confirm our values and to improve their accuracy. In
some cases, it turns out that it is not possible to find
lines of the secondary even with spectra of high S/N. It can therefore
be anticipated that cases with
small mass ratios $M_2/M_1$ are probably not uncommon.

Using the spectroscopic analysis and the partial light curve of KX~Vel,
we were able to obtain acceptable values of masses and radii. These 
parameters, however, better agree with a less luminous spectral 
classification than what is used now. New data, photometry, and
spectroscopy should be obtained to make these conclusions more reliable.

There are more than 100 binaries among southern OB objects; but
the fundamental parameters for only less than half of them are known.
Determining elements for more systems is necessary
to improve the statistics of mass ratios and other parameters of
OB-type binaries, and, of course, to give a better comparison of actually
observed quantities to the predictions of the theories of stellar formation,
structure, and evolution for this class of objects.

\begin{acknowledgement}
PM was supported by the grant P209/10/0715 of the Czech Science Foundation
and also by the research program MSM0021620860.
We kindly acknowledge helpful and valuable suggestions made by the referee
Dr. Otmar Stahl.
\end{acknowledgement}

\bibliographystyle{aa}

\end{document}